\documentclass[11pt]{article}
\usepackage{authblk}

\usepackage{enumitem}
\usepackage{graphicx,epsfig,amssymb,amstext,amsmath,amsthm}
\makeatletter
\newenvironment{tablehere}
{\def\@captype{table}}
{}

\makeatother

\usepackage{tikz}
\usepackage{graphics}
\usepackage{amsmath}
\usepackage{multicol}
\usepackage{amssymb}
\usepackage{float}
\usepackage{booktabs}
\newcommand{\bftab}{\fontseries{b}\selectfont}
\usepackage{multirow,tabularx}
\newcolumntype{Y}{>{\centering\arraybackslash}X}

\newcommand{\bo}{\boldsymbol}
\newcommand{\argmin}[1]{\underset{#1}{\operatorname{arg}\,\operatorname{min}}\;}
\usepackage{mathtools}
\DeclarePairedDelimiter{\ceil}{\lceil}{\rceil}
\DeclarePairedDelimiter{\floor}{\lfloor}{\rfloor}
\begin{document}

\title{Bayesian density regression for count data}
\author[1]{Charalampos Chanialidis}
\author[2]{Ludger Evers}
\author[3]{Tereza Neocleous}
\affil[1]{University of Glasgow, c.chanialidis1@research.gla.ac.uk}
\affil[2]{University of Glasgow, ludger.evers@glasgow.ac.uk}
\affil[3]{University of Glasgow, tereza.neocleous@glasgow.ac.uk}

\maketitle
\begin{abstract}
Despite the increasing popularity of quantile regression models for continuous responses, models for count data have so far received little attention. The main quantile regression technique for count data involves adding uniform random noise or ``jittering'', thus overcoming the problem that the conditional quantile function is not a continuous function of the parameters of interest. Although jittering allows estimating the conditional quantiles, it has the drawback that, for small values of the response variable $Y,$ the added noise can have a large influence on the estimated quantiles. In addition, quantile regression can lead to ``crossing'' quantiles. We propose a Bayesian Dirichlet process (DP)-based approach to quantile regression for count data. The approach is based on an adaptive DP mixture (DPM) of COM-Poisson regression models and determines the quantiles by estimating the density of the data, thus eliminating all the aforementioned problems. Taking advantage of the exchange algorithm, the proposed MCMC algorithm can be applied to distributions on which the likelihood can only be computed up to a normalising constant.
\end{abstract}







\section{Quantile regression}
Quantile regression was introduced as a nonparametric method for modelling a variable of interest as a function of covariates \cite{KB1978}. By estimating the conditional quantiles rather than the mean, it gives a more complete description of the conditional distribution of the response variable than least squares regression, and is especially relevant in certain types of applications. 

Consider a random variable $Y$ with cumulative distribution function $F(y)$. The $p$th quantile function of $Y$ is defined as \begin{equation}
Q(p)=\inf\{y \in \mathbb{R}:p\le F(y)\}\end{equation}
 and can be obtained by minimising the expected loss $E[\rho_p(Y-u)]$ with respect to $u$, where $\rho _{p}(y)=|y( p-I(y<0))|$. The $p$th sample quantile is obtained in a similar way by minimising $\sum_{i=1}^n \rho_p(y_i-u)$. 

Suppose that the $p$th conditional quantile function, $Q_Y(p|X=\boldsymbol{x})$, is a linear function of the predictors so that $ Q_Y(p|X=\bo{x})=X'\bo{\beta}_p.$
The parameter estimates $\bo{\hat{\beta}}_p$ are then obtained as 
\begin{equation}\bo{\hat{\beta}}_p=\argmin{\bo{\beta}_p \in \mathbb{R}^k}\sum_{i=1}^{n}\rho _{p}(Y-X'\bo{\beta}_p).\end{equation} 
A closed-form solution for this minimisation problem does not exist since the objective function is not differentiable at the origin, and it is solved using linear programming techniques \cite{B1998}. 



\subsection{Quantile regression for count data}
 The problem with applying quantile regression to count data is that the cumulative distribution function of the response variable is not continuous, resulting in quantiles that are not continuous, and which thus can not be expressed as a continuous function of the covariates. One way to overcome this problem is by adding uniform random noise (``jittering'') to the counts \cite{MS2002}. The general idea is to construct a continuous variable whose conditional quantiles have a one-to-one relationship with the conditional quantiles of the counts. Defining the new continuous variable $Z= Y + U$ where $Y$ is the count variable and $U$ is a uniform random variable in the interval $[0, 1)$, the conditional quantiles $Q_Z(p|X=\bo{x})= p + \exp(X'\bo{\beta}_p).$

The variable $Z$ is transformed in such a way that the new quantile function is linear in the parameters, i.e.$Q_{T(Z;p)}(p|X=\bo{x})=X'\bo{\beta}_p$ where

\begin{equation}
T(Z;p) = \left\{
\begin{array}{rl}\label{quantilestransform}
\log(Z-p) & \text{for } Z > p,\\
\log(\varsigma) & \text{for } Z \le p,
\end{array} \right.
\end{equation}
with $\varsigma$ being a small positive number. The parameters $\bo{\beta}_p$ are estimated by running a linear quantile regression of $T(Z;p)$ on $x$. 
Finally, the conditional quantiles of interest, $Q_Y(p|X=\bo{x})$ can be obtained from  the previous quantiles as
 \begin{equation}Q_Y(p|X=\bo{x})=\ceil*{Q_Z(p|X=\bo{x})-1}\end{equation} where $\ceil*{p}$ denotes the ceiling function which returns the smallest integer greater than, or equal to, $p$.  

While the jittering approach eliminates the problem of a discrete response distribution, for small values of the response variable $Y$, the mean and the variance in the transformed variable $Z$ will be mainly due to the added noise, resulting in poor estimates of the conditional quantiles  $Q_Y(p|X=\bo{x})$. As an example, when $Y=0$ the term $\log(Z-p)=\log(U-p)$ could go from $-\infty$ to $0$, simply due to the added noise.  In addition, quantile regression can suffer from the problem of crossing quantile curves, which is usually seen in sparse regions of the covariate space. This happens due to the fact that the conditional quantile curve for a given $X=\bo{x}$ will not be a monotonically increasing function of $p$. 

 Another approach would be to view the counts as ordinal variables with fixed thresholds and then model the new latent variable by an infinite mixture of normal densities \cite{KW2012}. Instead of using the aforementioned methods, we propose an adaptive Dirichlet process mixture approach which estimates the conditional density of the data. The approach is based on an adaptive Dirichlet Process mixture (DPM) of COM-Poisson regression models.
\section{COM-Poisson distribution}
 The COM-Poisson distribution \cite{CM1962,SMKBB2005} is a two-parameter generalisation of the Poisson distribution that allows for different levels of dispersion. The probability mass function of the COM-Poisson($\lambda$, $\nu$) distribution is
 \begin{equation}P(Y=y|\lambda, \nu)=\frac{\lambda^y}{(y!)^\nu}\frac{1}{Z(\lambda, \nu)} \ \  y=0, 1, 2, \ldots\end{equation}
where $Z(\lambda, \nu)=\displaystyle \sum_{j=0}^{\infty}\frac{\lambda^j}{(j!)^\nu}$ and $\lambda > 0$ and $\nu \ge 0$, where the normalisation constant does not have a closed form and has to be approximated numerically. The extra parameter $\nu$ allows the distribution to model under- ($\nu>1$) or over-dispersed ($\nu<1$) data, having the Poisson distribution as a special case ($\nu=1$). 

The above formulation of the COM-Poisson does not have a clear centering parameter since the parameter $\lambda$ is close to the mean only when $\nu$ takes values close to $1$, which makes it difficult to interpret for under- or over-dispersed data. Substituting the parameter $\lambda$ with $\mu=\lambda^{\frac{1}{\nu}}$, where $\floor{\mu}$ is the mode of the distribution
\begin{equation}
\mathbb{E}[Y]\approx \mu, \ \ \  \mathbb{V}[Y]\approx \frac{\mu}{\nu} \label{reform}
\end{equation}
and the new probability mass function is 
  \begin{equation}P(Y=y|\mu, \nu)=\left(\frac{\mu^y}{y!}\right)^\nu\frac{1}{Z(\mu, \nu)} \  \ y=0, 1, 2, \ldots \label{reformpmf}\end{equation}
where $Z(\mu, \nu)=\displaystyle \sum_{j=0}^{\infty}\left(\frac{\mu^j}{j!}\right)^\nu$.
\subsection{Mixtures of COM-Poisson distributions}
The COM-Poisson is flexible enough to approximate distributions with any kind of dispersion in contrast to a Poisson or a mixture of Poisson distributions which can only deal with overdispersion.

The two parameters of the COM-Poisson distribution allow it to have arbitrary (positive) mean and variance; one can obtain a point mass by letting the variance parameter $\nu$ tend to infinity. Thus one can show that mixtures of COM-Poisson distributions can provide an arbitrarily precise approximation to any discrete distribution with support $\mathbb{N}_0$, which is why COM-Poisson distributions are used by our method. All other generalisations of the Poisson distribution we are aware of do not have this property. 

\subsection{COM-Poisson regression}
A regression model can be defined based on \eqref{reformpmf}, in which both the mean and the variance parameter are modelled as a function of covariates:
 \begin{align}
 \log{\bo{\mu_i}}&= \bo{x_i}^\intercal\bo{\beta}\\
 \log{\bo{\nu_i}}&= \bo{x_i}^\intercal\bo{c}
  \end{align}
where $Y$ is the response variable being modelled, and $\bo{\beta}, \bo{c}$ are the regression coefficients for the centering link function and the shape link function respectively. The parameters in this formulation have a direct link to either the mean or the variance, providing insight into the behaviour of the response variable. Notably,
\begin{equation}
\mathbb{E}[Y_i]\approx \exp(\bo{x}_i'\bo{\beta}), \ \ \ \mathbb{V}[Y]\approx \frac{\exp(\bo{x}_i'\bo{\beta})}{\exp(\bo{x}_i'\bo{c})} =\exp(\bo{x}_i'(\bo{\beta}-\bo{c})).
\end{equation}
The calculation of the normalisation constant of the COM-Poisson distribution is the computationally most expensive part of the proposed regression model. It can be seen, in the next subsection, that this calculation is redundant.

\subsection{Exchange algorithm}
Any probability density function $p(y|\theta)$ can be written as
\begin{equation}p(y|\theta)=\frac{q_\theta(y)}{Z(\theta)}\end{equation} where $q_\theta(y)$ is the unnormalised density and the normalising constant $Z(\theta)=\int p(y,\theta) \, \mathrm{d}y $  is unknown. In this case the acceptance ratio of the Metropolis-Hastings algorithm is
\begin{align}
\alpha&=\min \left(1,\frac{q_{\theta^*}(y) \pi(\theta^*) Z(\theta)h(\theta|\theta^*)}{q_\theta(y) \pi(\theta) Z(\theta^*)h(\theta ^*|\theta)}\right)\label{defaultmh}
\end{align}
where $\pi (\theta)$ is the prior distribution of $\theta$. The acceptance ratio in \eqref{defaultmh} involves computing unknown normalising constants. Introducing auxiliary variables $\theta^*, y^*$ and sampling from an augmented distribution
\begin{equation}\pi (\theta ^*,y^*,\theta|y) \propto p(y|\theta)\pi (\theta) p(y^*|\theta ^*) h(\theta ^*|\theta) \end{equation}
results in
\begin{align}
\alpha&=\min \left(1,\frac{p(y|\theta^*)\pi (\theta^*) p(y^*|\theta) h(\theta|\theta^*)}{p(y|\theta)\pi (\theta) p(y^*|\theta ^*) h(\theta ^*|\theta)}\right)\\
&=\min \left(1,\frac{q_\theta(y^*)\pi (\theta^*) h(\theta|\theta^*)q_{\theta^*}(y) Z(\theta) Z(\theta^*) }{q_\theta(y) \pi (\theta) h(\theta ^*|\theta) q_{\theta^*}(y^*) Z(\theta^*)  Z(\theta) }\right)\\
&=\min \left(1,\frac{q_\theta(y^*)\pi (\theta^*) q_{\theta^*}(y) }{q_\theta(y) \pi (\theta) q_{\theta^*}(y^*) }\right) \label{exchangemh}
\end{align}
where the normalising constants cancel out and $h()$ is a symmetric distribution \cite{MGM06,MPRB2006}. In order to be able to use this algorithm one has to be able to sample from from the unnormalised density which in the case of the COM-Poisson distribution can be done efficiently using rejection sampling. 

Updating the parameter $\mu$ of the COM-Poisson we have $\theta=(\mu,\nu)$ and $\theta^*=(\mu^*,\nu)$
where $\mu^*$ follows a Normal distribution centered at $\mu$ and 
\begin{align}
q_\theta(y^*)&=\left(\frac{\mu_i^{y_i^*}}{y_i^*!}\right)^{\nu_i} & q_{\theta^*}(y)&=\left(\frac{(\mu_i^*)^{y_i}}{y_i!}\right)^{\nu_i}\\
q_\theta(y)&=\left(\frac{\mu_i^{y_i}}{y_i!}\right)^{\nu_i}&q_{\theta^*}(y^*)&=\left(\frac{(\mu_i^*)^{y_i^*}}{y_i^*!}\right)^{\nu_i}
\end{align}
Likewise for updating the parameter $\nu$.

\section{Bayesian density regression}
Density regression is similar to quantile regression in that it allows flexible modelling of the response variable $Y$ given the covariates $\boldsymbol{x}=(x_1, \ldots, x_p)'$. Features (mean, quantiles, spread) of the conditional distribution of the response variable, vary with  $\boldsymbol{x}$, so, depending on the predictor values, features of the conditional distribution can change in a different way than the population mean. The difference between density regression and quantile regression is that density regression models the probability density function or probability mass function rather than directly modelling the quantiles. 

\subsection{Bayesian density regression for count data}
This paper focuses on the following mixture of regression models:
\begin{equation}f(y_i|\boldsymbol{x}_i)=\int f(y_i|\boldsymbol{x}_i,\phi_i)G_{\boldsymbol{x}_i}(\mathrm{d}\phi_i)\end{equation}
where \begin{equation}f(y_i|\boldsymbol{x}_i,\phi_i)=\text{COM-P}(y_i;\exp(\boldsymbol{x}_i'\boldsymbol{b}_i),\exp(\boldsymbol{x}_i'\boldsymbol{c}_i))\end{equation}
the conditional density of the response variable given the covariates is expressed as a mixture of COM-Poisson regression models  with $\phi_i=(\boldsymbol{b}_i,\boldsymbol{c}_i)$ and $G_{\boldsymbol{x}_i}$ is an unknown mixture distribution that changes according to the location of $\boldsymbol{x}_i$.

\subsection{MCMC algorithm}
Let $\boldsymbol{\theta}=(\theta_1, \ldots, \theta_k)'$ denote the $k\le n$ distinct values of $\phi$ and let $\boldsymbol{S}=(S_1, \ldots, S_n)'$ be a vector of indicators denoting the global configuration of subjects to distinct values $\boldsymbol{\theta}$, with $S_i=h$ indexing the location of the $i$th subject within the $\boldsymbol{\theta}$. In addition, let $\boldsymbol{C}=(C_1, \ldots, C_k)'$  with $C_h=j$ denoting that $\theta_h$ is an atom from the basis distribution, $G_{\boldsymbol{x}_j}^*$. Hence $C_{S_i}=Z_i=j$ denotes that subject $i$ is drawn from the $j$th basis distribution.

Excluding the $i$th subject, $\boldsymbol{\theta}^{(i)}=\boldsymbol{\theta}\backslash \{\phi_i\}$ denotes the $k^{(i)}$ distinct values of $\boldsymbol{\phi}^{(i)}=\boldsymbol{\phi}\backslash \{\phi_i\}$, $\boldsymbol{S}^{(i)}$ denotes the configuration of subjects $\{1, \ldots, n\}\backslash \{i\}$ to these values and  
$\boldsymbol{C}^{(i)}$ indexes the DP component numbers for the elements of  $\boldsymbol{\theta}^{(i)}$.

Grouping the subjects in the same cluster and updating the prior with the likelihood for the data  $\boldsymbol{y}$, we obtain the conditional posterior
\begin{equation}(\phi_i|\boldsymbol{S}^{(i)},\boldsymbol{C}^{(i)},\boldsymbol{\theta}^{(i)} ,\boldsymbol{X},a) \sim q_{i0}G_{i,0} + \sum_{h=1}^{k^{(i)}} q_{ih}\delta_{\theta_h^{(i)}},\end{equation}
where $G_{i,0}(\phi)$ is the posterior obtained by updating the prior $G_0(\phi)$ and the likelihood $f(y_i|\boldsymbol{x}_i,\phi)$:
\begin{align}
  G_{i,0}(\phi)&= \frac{G_0(\phi)f(y_i|\boldsymbol{x}_i,\phi)}{h_i(y_i|\boldsymbol{x}_i)}, & &\\
  q_{i0}&= cw_{i0}h_i(y_i|\boldsymbol{x}_i),& q_{ih}&=cw_{ih}f(y_i|\boldsymbol{x}_i,\theta_h),\\
  w_{i0}&=\sum_{j=1}^n \frac{ab_{ij}}{a + \sum_{l\ne i}\boldsymbol{1}(C_{S_l^{(i)}}^{(i)}=j)}, &w_{ih}&=\frac{b_{i,C_h^{(i)}}\sum_{m\ne i}\boldsymbol{1}(S_m^{(i)}=h)}{a+ \sum_{l\ne i}\boldsymbol{1}(C_{S_l^{(i)}}^{(i)}=C_h)},
\end{align}

where $b_{ij}$ are weights that depend on the distance between subjects' predictor values, $c$ is a normalising constant and $h=1, \ldots , k^{(i)}$. Since there is no closed form expression for the posterior distribution, approximation of the probability $q_{i0}=cw_{i0}h_i(y_i|\boldsymbol{x}_i)$ is difficult. 

We overcome this problem by bridging: i) an MCMC algorithm for sampling from the posterior distribution of a Dirichlet process model, with a non-conjugate prior, found in \cite{N2000}; ii) the MCMC algorithm found in \cite{DPP2007}; and iii) a variation of the MCMC exchange algorithm. 



 The MCMC algorithm alternates between the following steps:
\begin{description}
\item[Step 1:] Update $S_i$ for $i=1, \ldots, n,$ by proposing, from the conditional prior,  a move to a new cluster or an already existing cluster with probabilities proportional to $w_{i0}$ and $w_{ih}$ for $h=1, \ldots,k^{(i)}.$ 
\begin{enumerate}
\item[a)] If the proposed move is to go to a new cluster we draw parameters ($\mu_0,\nu_o$) for that cluster from $G_0$ and at the same time sample an observation $y^*$ from the COM-Poisson($\mu_0,\nu_0$). The acceptance ratio of the Metropolis-Hastings algorithm is 
\begin{equation}\min \left(1,\frac{q_\theta(y^*)q_{\theta^*}(y) }{q_\theta(y) q_{\theta^*}(y^*) }\right) \label{exchangemh2}
\end{equation}
If the proposal is accepted,  $C_{S_i} \sim$ multinomial $(\{1, \ldots, n\},\boldsymbol{b}_i)$.
\item[b)] If the proposed move is to an already existing cluster $h$, we sample an observation $y^*$ from the COM-Poisson($\mu_h,\nu_h$) and accept with the same probability as in \eqref{exchangemh2}. If the proposal is accepted $C_{S_i}=C_h$.
 \end{enumerate}

\item[Step 2:] Update the parameters $\theta_h$, for $h=1, \ldots, k$ by sampling from the conditional posterior distribution
\begin{equation}(\theta_h|\boldsymbol{S},\boldsymbol{C},\boldsymbol{\theta}^{(h)} ,k,\boldsymbol{y},\boldsymbol{X}) \sim \prod_{i:S_i=h} f(y_i|\boldsymbol{x}_i,\theta_h) \} G_0(\theta_h),\end{equation}
using the  Metropolis-Hasting algorithm with acceptance probability as in \eqref{exchangemh}.
\item[Step 3:] Update $C_h$, for $h=1, \ldots, k$, by sampling from the multinomial conditional with 
\begin{equation}(C_h|\boldsymbol{S},\boldsymbol{C}^{(h)},\boldsymbol{\theta} ,k,\boldsymbol{y},\boldsymbol{X}) \sim \frac{\prod_{i:S_i=h} b_{ij}}{\sum_{l=1}^n \prod_{i:S_i=h} b_{il}}, \ j=1, \ldots, n\end{equation} and location weights $\gamma_j$ for $j=1, 2, \ldots, n$, using an approach used in \cite{DS2005}. 
\end{description}

\section{Simulations and application}
We consider two simulated data sets to compare the proposed discrete Bayesian density regression method to the ``jittering'' method. These are
\begin{align}
Y_i|X_i=x_i&\sim \text{Binomial}(10,0.3x_i)\\
Y_i|X_i=x_i&\sim 0.4\text{Pois}(\exp(1+x_i)) + 0.2\text{Binomial}(10,1-x_i) +0.4 \text{Geom}(0.2)
\end{align} 
 where $x_i \sim \text{Unif}(0,1)$. Table~\eqref{difference} shows the absolute mean errors obtained using both methods. If $q_p$ is the true conditional quantile when $x=p$ and $\hat{q_p}$ is the estimated conditional quantile, the mean absolute error is defined as $\mathbb{E}[|q_p - \hat{q_p}|]$. The discrete Bayesian density regression (BDR) estimates outperform the ``jittering'' method and in almost all cases the ``jittering'' method leads to crossing quantiles (except when $n=500$).

\begin{tablehere}
\centering
\tabcolsep=0.05cm
\scalebox{0.9}{
\begin{tabular}{lcccccc}
Method& \multicolumn{6}{c}{\centering Number of Observations} \\ \hline\noalign{\smallskip}
&\multicolumn{3}{c}{\centering Binomial}&\multicolumn{3}{c}{\centering Mixture}  \\
  &\multicolumn{1}{c}{20} & \multicolumn{1}{c}{100} & \multicolumn{1}{c}{500}&\multicolumn{1}{c}{20} & \multicolumn{1}{c}{100} & \multicolumn{1}{c}{500}\\ \hline\noalign{\smallskip}  
Density Regression & 0.5576 &   \bftab{0.2820} & \bftab{0.2421} & \bftab{0.7435} &   \bftab{0.5833} & \bftab{0.3589}\\
Jittering (linear) & \bftab{0.5256}  & 0.8461  & 0.4765 & 1.1923 & 0.6666  & 0.4294\\
Jittering (splines) &0.7820 &0.5128 & 0.3020 &1.9487 & 0.8269 & 0.3910\\
\noalign{\smallskip}\hline\noalign{\smallskip}
\end{tabular}}
\caption{Mean absolute error obtained using the different density/quantile regression methods. 
} \label{difference}
\end{tablehere}

\begin{figure}
\centering
\includegraphics[height=7.2cm]{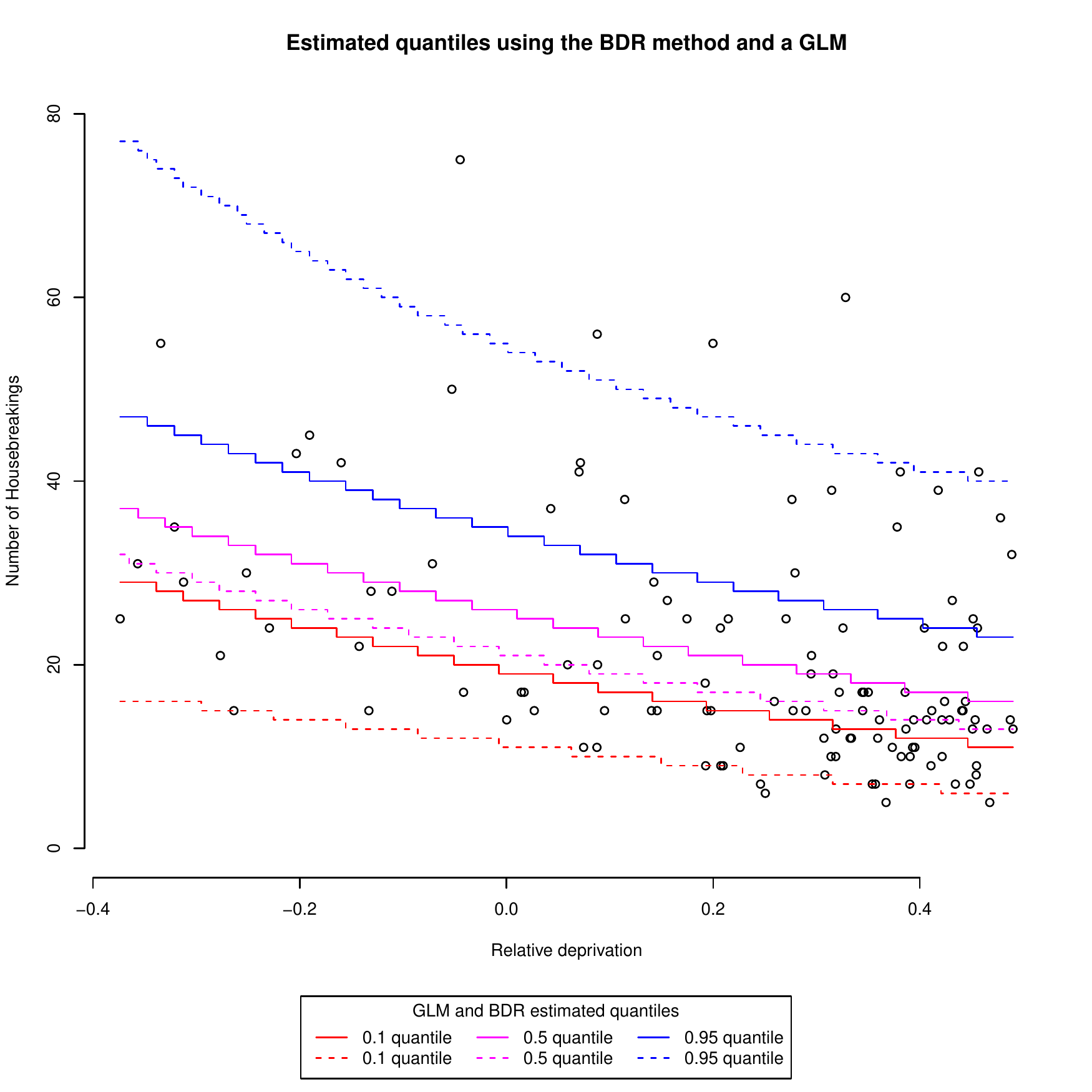}
\caption{Estimated quantiles for housebreaking data, using discrete Bayesian density regression (dashed lines) and derived from a Poisson model.}\label{fig:glm2}
\end{figure}

We apply the discrete density regression technique to data on housebreakings in Greater Glasgow (Scotland). The data consist of the number of housebreakings in each of the 127 intermediate geographies in Greater Glasgow in 2010.  We aim to relate the number of housebreakings to the deprivation score of the intermediate geography area, as measured by the Scottish Index of Multiple Deprivation (SIMD). The deprivation score is standardised by considering the difference of each intermediate geography's deprivation from the average deprivation in Greater Glasgow e.g. low values relate to affluent areas, large values to deprived areas. 
The solid and dashed lines in figure~\ref{fig:glm2} show the quantiles (for $p=0.1, 0.5, 0.95$) obtained for the standard Poisson regression model and the COM-Poisson model respectively. The first model is not able to capture the overdispersion of the data, nor the skewness of the distribution. 

\section{Conclusions and further research}
In this manuscript we have proposed a novel Bayesian density regression technique for discrete data which is based on mixing COM-Poisson distributions. The new method takes advantage of the exchange algorithm and updates the cluster allocations by drawing a new allocation for an auxiliary observation and then accepting or rejecting it. As a result the MCMC samples from the target distribution without the need to estimate the normalisation constant of the likelihood. The method overcomes the two main drawbacks of the ``jittering'' method for discrete quantile regression, namely that it does not require the addition of artificial additional noise and that it does not suffer from the problem of crossing quantiles. We have illustrated the method in a real world application as well as simulated examples in which our method compared favourably to the ``jittering'' method. Further research efforts will be devoted in improving the computational speed and efficiency of the MCMC algorithm to make it an even more attractive alternative to ``jittering''.

\end{document}